\documentclass[a4paper,11pt]{article}
\usepackage[utf8]{inputenc}
\usepackage{graphicx}
\usepackage{amsmath}
\pdfoutput=1
\usepackage{slashed,mathtools}
\usepackage{jheppub}
\usepackage{subfig}
\usepackage{tikz}
\usepackage{multirow}
\allowdisplaybreaks

\title{Resonant Assisted Annihilation}

\author[a]{Tarak Nath Maity,}
\author[a,b]{Tirtha Sankar Ray}
\affiliation[a]{Department of Physics, Indian Institute of Technology Kharagpur, Kharagpur 721302, India}
\affiliation[b]{Centre for Theoretical Studies, Indian Institute of Technology Kharagpur, Kharagpur 721302, India}
\emailAdd{tarak.maity.physics@gmail.com}
\emailAdd{tirthasankar.ray@gmail.com}

\abstract{Assisted annihilation is a novel mechanism to generate viable sub-GeV thermal dark matter, where a pair of stable dark matter annihilates with an assister to Standard Model states. Typically such $3 \to 2$ annihilation topologies  are flux suppressed compared to  $2 \to 2$ processes. In this paper, we explore the possibility of a resonant $3 \to 2$ assisted annihilation  dominantly driving the freeze-out of dark matter. We demonstrate that in a simple multipartite scalar extension of the Standard Model this can be realized in certain regions of parameter space to provide viable dark matter relic density, in agreement with observation. We demonstrate that  for photophilic assisters parts of the parameter space are already constrained by indirect detection experiments and the measurements of CMB anisotropies  while substantial regions remain beyond the present limit.   
}


\begin{document}

\maketitle
\flushbottom

\section{Introduction}
\label{sec:intro}
Sub-GeV dark matter (DM) scenarios have received recent attention from various quarters spurred by cosmological observations  at the galactic scale in the context of structure formation \cite{ Spergel:1999mh, Nakama:2017ohe, Bullock:2017xww}.  Independently a lot of effort have been made to update the direct detection  experiments to target sub-GeV weakly interacting DM \cite{Crisler:2018gci, Abramoff:2019dfb, Essig:2011nj,  Essig:2015cda, Lee:2015qva, Hochberg:2016sqx, Kurinsky:2019pgb, Hochberg:2015pha, Hochberg:2016ajh, Hochberg:2019cyy, Dror:2019onn}.  These have led to a renewed interest in motivated model building for sub-GeV DM beyond the standard model \cite{Battaglieri:2017aum}. Generalizing the standard paradigm of $2\to 2$ annihilation to  $N(>2) \to 2$ topologies naturally lead to light DM in sub-GeV domain \cite{Dolgov:1980uu, Dolgov:2017ujf, Hochberg:2014dra, Hochberg:2014kqa, Dey:2016qgf, Bernal:2015xba, Lee:2015uva, Choi:2015bya, Bernal:2015bla, Hochberg:2015vrg, Kuflik:2015isi, Choi:2016tkj, Choi:2016hid, Bernal:2017mqb, Ho:2017fte, Cline:2017tka, Choi:2017zww, Kuflik:2017iqs, Hochberg:2018rjs, Hochberg:2018vdo, Bhattacharya:2019mmy, Chauhan:2017eck, Choi:2017mkk}. An effort in this direction  is the so called \textit{assisted annihilation} framework that was introduced in \cite{Dey:2016qgf}. Minimal version of this class of  models has a  sub-GeV  stable thermal DM  state along with assisters that can promptly decay to SM.  By construction, the  annihilation to  SM states in the early universe is dominated by a $N\to 2$ topology where a pair of DM particles annihilate with one or more  assisters in the initial state.  Interestingly since the assisters are not charged under the same stabilizing  symmetry as the DM it can in principle be lighter  leading to a Boltzmann boost to the annihilation process \cite{Dey:2018yjt}. There are non-trivial effect of these new light states on the cosmology of the early Universe. Some of the relevant constraints on this framework from Big Bang Nucleosynthesis (BBN) and Cosmic Microwave Background (CMB) have been explored in  \cite{Dey:2018yjt}. 
Additionally this class of models remain insulated  from the present and  proposed direct detection experiments  due to additional flux suppression.

From the point of view of  model building  the challenge is to have  the flux suppressed  $3\to 2$ channel dominate over possible $2 \to 2 $ processes. A possibility of eliminating the $2 \to 2 $ channel by a combination of kinematic phase space and Boltzmann suppression was  presented in  \cite{Dey:2018yjt}. This required augmentation of the minimal setup to include  a heavy mediator in addition to the DM and assister states.  The object was to suppress the associated  $2 \to 2$ process without tuning couplings. Keeping within this setup, in this paper we take the complimentary view of boosting the assisted annihilation process by tuning it near a resonance peak. For universal couplings the resonant $s$-channel mediated  $3 \to 2$ can easily dominate over $2 \to 2 $ processes.  We find that for such scenarios it is easy to have an assisted  annihilation dominated freeze-out of DM that saturates the observed relic abundance limits with perturbative couplings.

In  this paper, we present a simple scalar model of assisted annihilation  containing a scalar $\mathbb{Z}_2$ odd DM, a photophilic scalar assister and a scalar mediator. We explore the region of parameter space where the assisted annihilation is near resonance. We demonstrate that within this framework  it is easy to match the observed relic density of DM with perturbative couplings.  We briefly comment on the possibility of probing a part of the relic density allowed parameter space  of this framework using  indirect detection  and beam dump experiments.

The paper is organized as follows. In section \ref{sec:model} we present the details of the minimal scalar model for  resonant assisted annihilation. We make a systematic study of the relic density of DM within this framework in section \ref{sec:relic}. We discuss  the possibility of  exploring this framework in indirect detection and beam dump experiments in section \ref{sec:pheno} before concluding.

\section{Minimal Model for Resonant \textit{Assisted Annihilation}}
\label{sec:model}
The minimal real scalar model for resonant assisted annihilation contains three real scalar fields viz. a stable  DM ($\phi$), an assister $(A)$  and  heavy mediator $(S)$ which are all singlet under the SM gauge symmetries. The stability of the DM is ensured by assigning an odd charge to it under a discrete $\mathbb{Z}_2$ symmetry, while both the assisters and  mediator can promptly decay to SM states which are all even under the  same $\mathbb{Z}_2$.   The  assister  and the mediator also  double up  as a portal to the visible sector keeping DM in thermal equilibrium before freeze-out. The most general  scalar potential for the dark sector and a photophilic portal coupling to the visible sector consistent with aforementioned charge assignments can be written as
\begin{equation} \label{eq:pot1}
\begin{split}
\mathcal{L}_{\mbox{dark}} &=  \frac{1}{2} m_{\phi}^2 \, \phi^2 \, + \, \frac{1}{2}  m_{A}^2 \, A^2 \, + \frac{1}{2} \, m_{S}^2 \, S^2     
                            \\ &+ \, \frac{\lambda_1}{4} \, \phi^2 A^2 \, + \,  
                            \frac{\lambda_2}{4} \, \phi^2 S^2 \, 
                            + \, \frac{\lambda_3}{2} \, \phi^2  AS \,  + \,   
                             \frac{\lambda_4}{4}\, A^2 S^2 \,  \\&+ \, \frac{\lambda_5}
                             {6}\, A^3 S \, + \,  \frac{\lambda_6}{6} \, S^3 A  + \frac{\mu_1}{2} \phi^2 A + \frac{\mu_2}{2} \phi^2 S \\
                            &+ \frac{\mu_3}{6} A^3 \, + \, \frac{\mu_4}{6} S^3 \, + \, \frac{\mu_5}{2} A^2 S \, + \, \frac{\mu_6}{2} \, S^2 A \, \\
                            \mathcal{L}_{\mbox{portal}} &= \, c^{a}_{\gamma} AF^{\mu \nu}F_{\mu \nu}  +c^{s}_{\gamma}SF^{\mu \nu}F_{\mu \nu}, 
\end{split}
\end{equation}
where $F^{\mu \nu}$ is the standard electromagnetic field strength tensor and $c^{a}_{\gamma}, c^{s}_{\gamma} $ have mass dimension of minus one.  The non-renormalizable portal coupling  of the dark sector to the SM represents a special choice which enables us  to  extract the  most interesting  phenomenological implications  of the framework. Generalizations are straightforward and   do not affect the resonant assisted annihilation  driven freeze-out of DM discussed in the next section. 
\begin{figure}[th]
\begin{center}
\subfloat[\label{sf:2phito2A}]{
\includegraphics[scale=0.20]{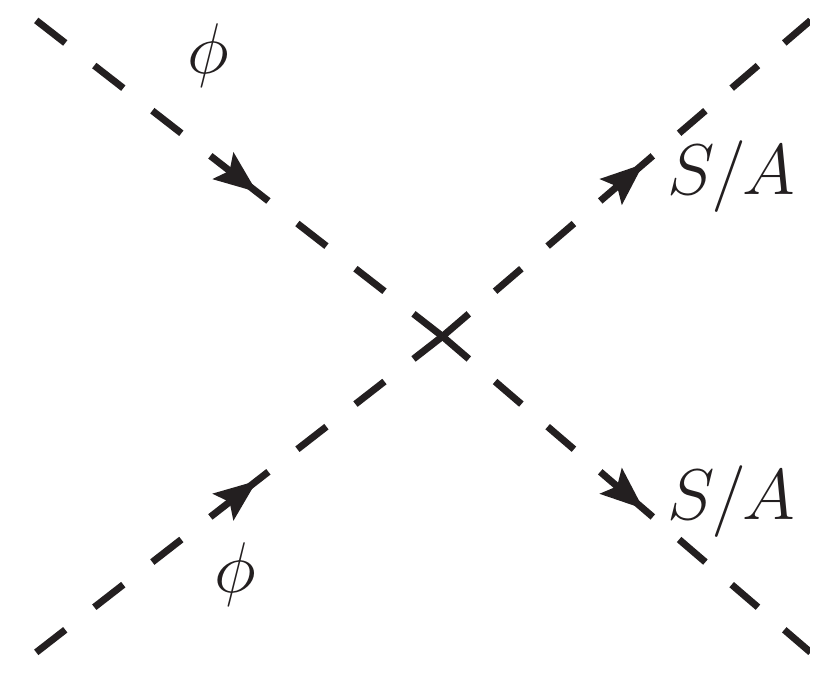}~~~~
\includegraphics[scale=0.20]{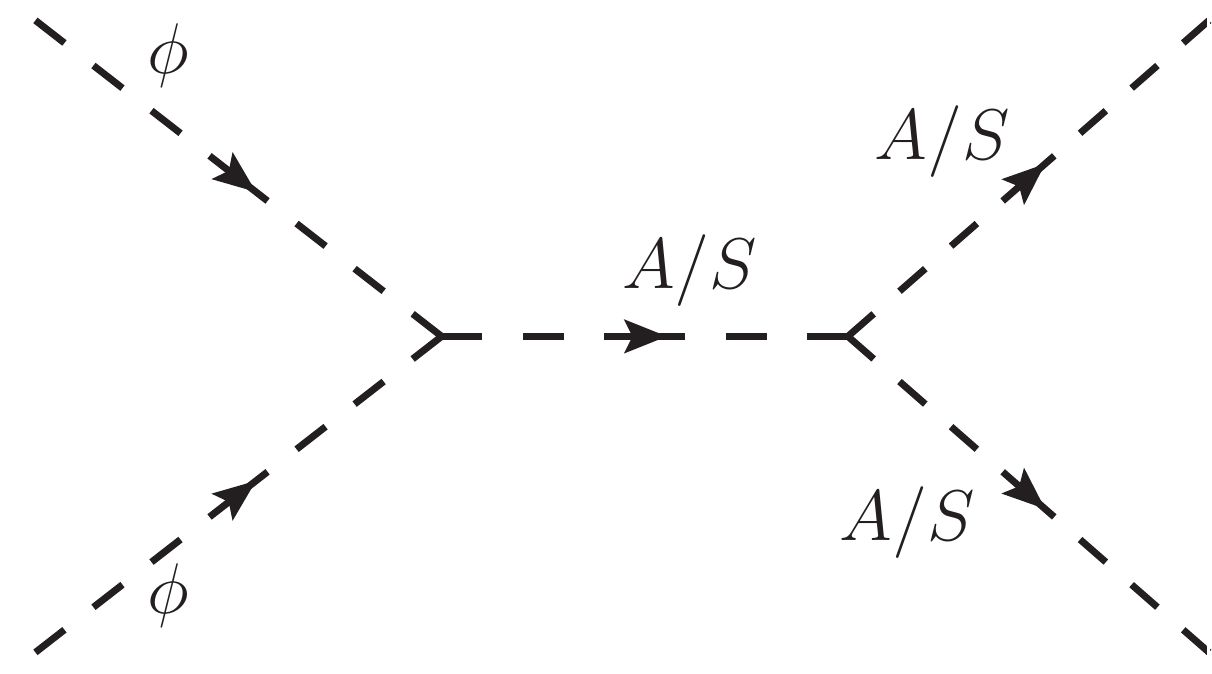}}~~~~
\subfloat[\label{sf:2phito2g}]{
\includegraphics[scale=0.18]{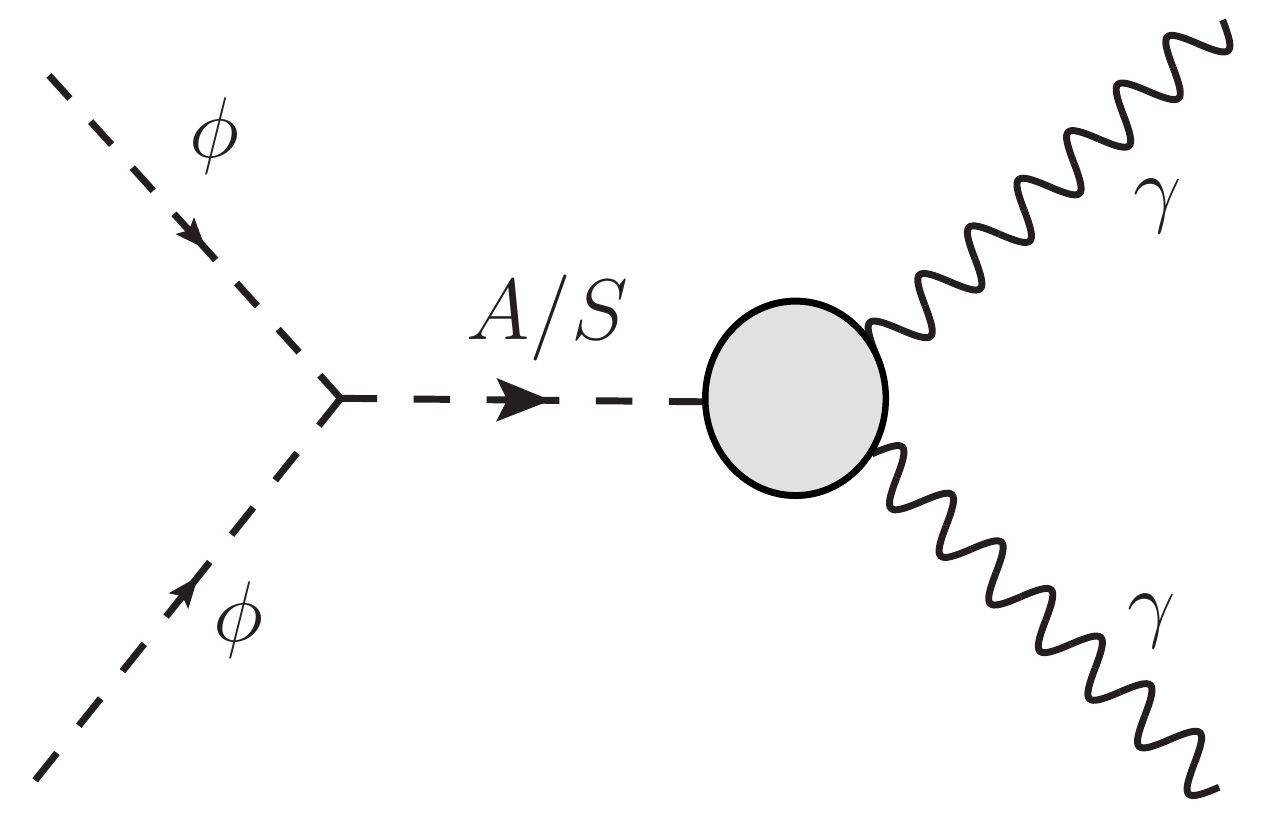}}~~~~
\subfloat[\label{sf:2phiAtoAA}]{
\includegraphics[scale=0.20]{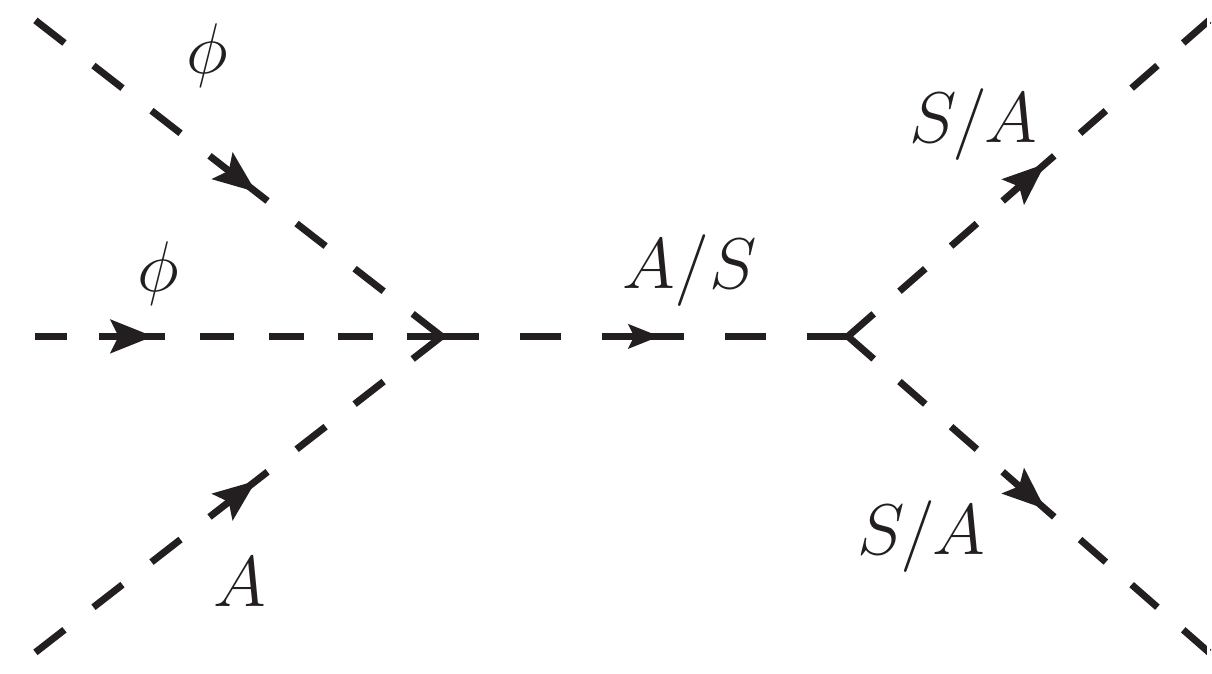}}~~~~
\subfloat[\label{sf:2phiAto2g}]{
\includegraphics[scale=0.20]{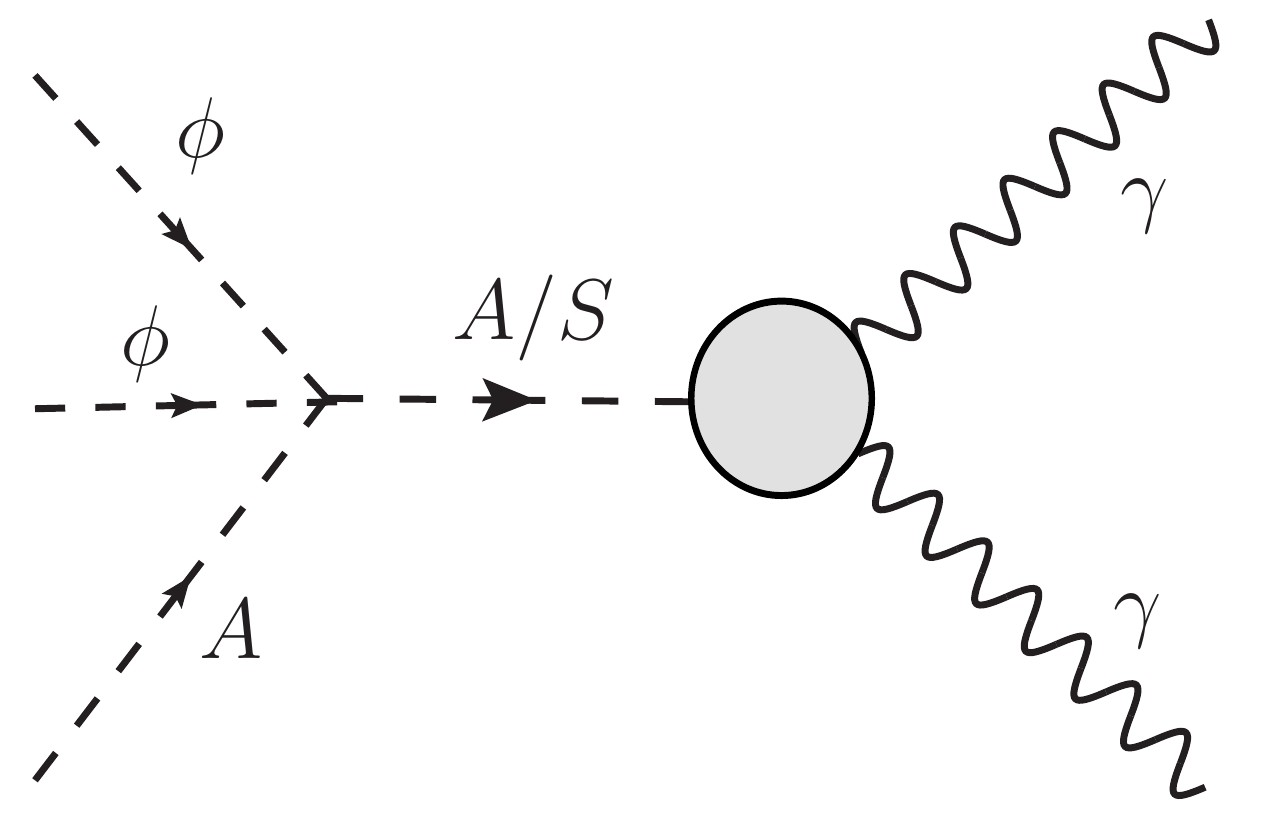}}
\caption{Feynman diagram of the relevant $2 \to 2$  and $3 \to 2$ processes.}
\label{fig:fd}
\end{center}
\end{figure}
As can be easily read out from the interactions  given in equation \eqref{eq:pot1}   one of the annihilation channel  for the DM proceed through the novel $3 \to 2 $ topology given by  $\phi \phi A \rightarrow S \rightarrow A A / \gamma \gamma .$ Ordinarily this will be overwhelmed by a host of $2\to2$ processes like $\phi \phi \rightarrow SS / AA /AS$ etc.  However in certain regions of the parameter space  where the masses are tuned to put the assisted annihilation process on $s$-channel resonance this will dominantly drive the freeze-out of DM.  In the next section, we will explore this possibility of resonant assisted annihilation  setting the DM relic density.  Subsequently, we will explore the phenomenological consequences of this framework.

\section{Relic Density}
\label{sec:relic}
In this section we will focus on the region of parameter space where the resonant assisted annihilation processes (shown in figure \ref{sf:2phiAtoAA} and \ref{sf:2phiAto2g}) set the required relic density of DM \cite{Aghanim:2018eyx}. To keep the discussion tractable  we will further assume that all the $\lambda_i$, $\mu_i$ and $c^{i}_{\gamma}$ of equation \eqref{eq:pot1} are universal and  equal to  $\lambda$, $\mu$ and $c_{\gamma}$ respectively. However, to have an handle on the relative strength of the  $2 \to 2$ and $3 \to 2$ processes we keep $\lambda_3$ as an independent coupling. We call this four parameter scenario as the Benchmark Model. Admittedly, this requires a tuning of masses of the dark sector states of the form, $(2m_\phi + m_A ) \sim m_S.$  The relevant Boltzmann equation is given by,
\begin{subequations}
\begin{align}
\label{seq:boltzYY}
 \frac{dY_{\phi}}{dx} &= -\frac{s^{2} g_{*}}{x H}
                        N_{\rm Bolt}
                        \langle \sigma v^{2}\rangle_{3 
                        \to 2}
                        \left[Y_{\phi}^{2}
                        Y_{\phi}^{\rm eq}
                        - \left(Y_{\phi}^{\rm 
                        eq}\right)^{3}
                        \right] -\frac{s g_{*}}{x H}                       
                      \langle \sigma v \rangle_{2 \to 2} 
                        \left[Y_{\phi}^{2}
                        - \left(Y_{\phi}^{\rm eq}\right)^{2}
                        \right] \\
\label{seq:nbolt}
N_{\rm Bolt} &= e^{x(1-\epsilon)}\epsilon^{3/2},~~g_{*} = 1 + \frac{1}{3} \frac{d(\text{ln}~g_s)}{d(\text{ln}~T)},       
\end{align}
\label{eq:boltzy}
\end{subequations}
where $x=m_{\phi}/T$, $\epsilon=m_{A}/m_{\phi}$, entropy density $s=2 \pi^2 g_s T^3/45$, Hubble constant $H= \sqrt{\pi^2 g_{\rho}/90}\left(T^2/M_{\rm Pl}\right)$, $g_s$ and $g_{\rho}$ are the effective number of relativistic degrees of freedom corresponding to entropy and energy density respectively. For the temperature dependence of $g_{\rho}$, $g_{s}$ and $g_{*}$ we have followed \cite{Drees:2015exa}. The thermally averaged cross section $\langle \sigma v^{2}\rangle_{3 \to 2}$ includes all $3 \to 2$  processes while  $\langle \sigma v \rangle_{2 \to 2}$  quantify the sub-dominant  $2 \to 2$ annihilation cross sections\footnote{In principle to obtain the correct relic density full set of coupled Boltzmann equations involving the DM, assisters and mediator should be considered. However, at resonance, equation \eqref{eq:boltzy} reproduces the results adequately.}. 
Note that, for $\epsilon > 1$ there will be a Boltzmann suppression of the initial state assister flux while for $\epsilon < 1$ there is an enhancement of the  effective cross section for similar reasons. This is a novel feature of the assisted annihilation framework that should be contrasted with the usual co-annihilation scenario, where by construction a Boltzmann suppression is obtained depending on the mass-splitting of the co-annihilating states \cite{Dey:2018yjt}.
\begin{figure}[t]
\begin{center}
\includegraphics[scale=0.3]{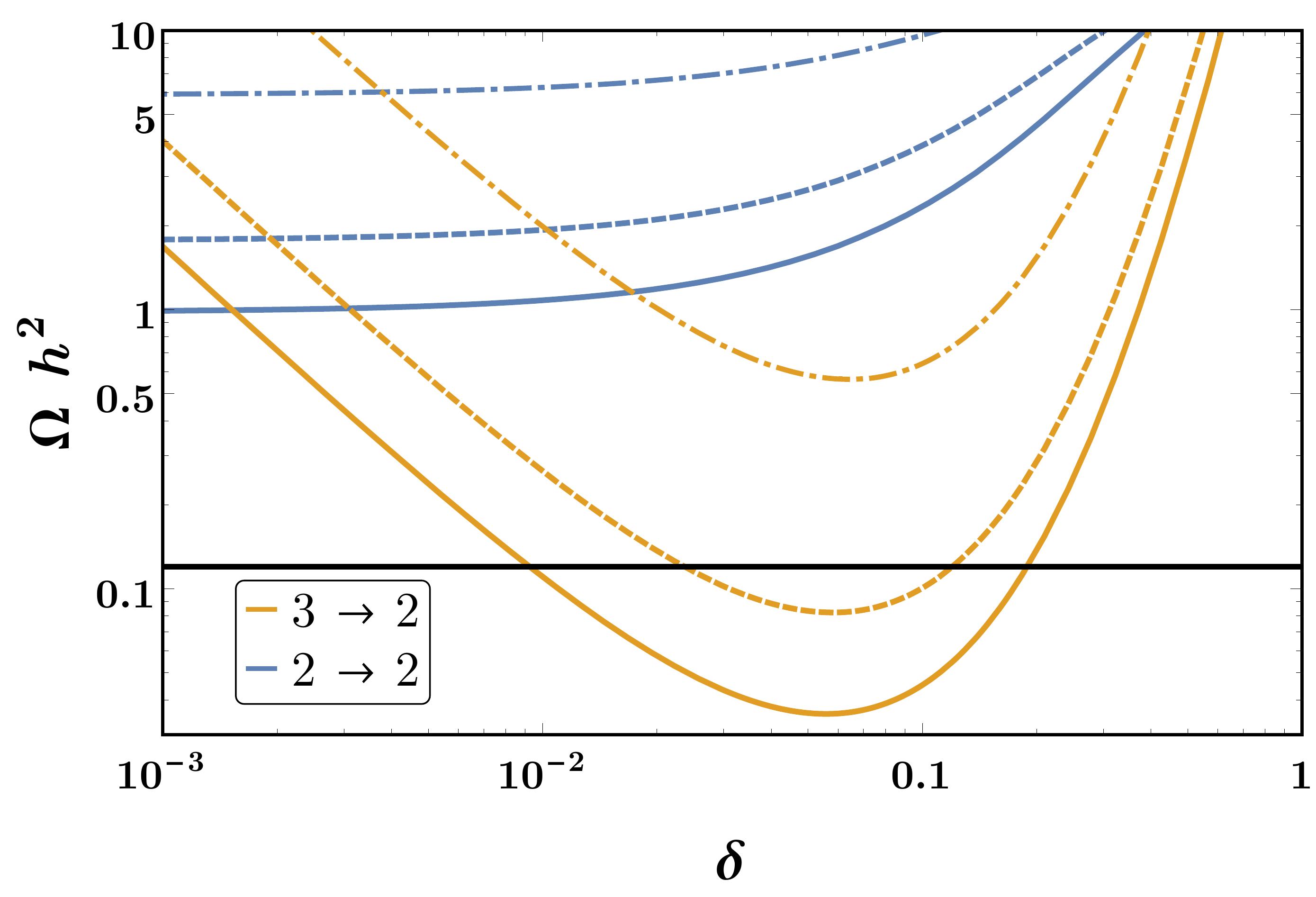}
\caption{Relic density as a function of $ \delta$, where $\delta$ has been defined in equation \eqref{eq:delta}. The light blue and light orange lines show $\Omega h^2$ estimated using $2 \to 2$ and $3 \to 2$ channels for different choices of $\lambda_3$. For the solid, dashed and dot dashed lines $\lambda_3$ has been fixed to $1,~0.5,~\rm and~ 0.1$ respectively.  We set $m_{\phi} = 200$ MeV, $m_{A} = 100$ MeV, $\lambda =10^{-5}$, $\mu/m_{\phi}=10^{-5}$  and  $c_{\gamma}=10^{-11}$  MeV$^{-1}$.}
\label{fig:resonance}
\end{center}
\end{figure} 
The relevant Feynman diagrams of $2 \to 2$ processes are shown in figure \ref{sf:2phito2A} and \ref{sf:2phito2g} while the Feynman diagrams of $3 \to 2$ assisted annihilation processes are shown in figure  \ref{sf:2phiAtoAA} and \ref{sf:2phiAto2g}. Note that the cross section of both $3 \to 2$ and $2 \to 2$ processes depend on $\lambda,~\lambda_3,~\mu$ and $c_{\gamma}$. In spite of the strong constraint on $c_{\gamma}$ from fixed target experiments \cite{Aloni:2019ruo} with  $\lambda$ and $\mu/m_{\phi} \sim \mathcal{O}(1)$ the $2 \to 2$ processes in general  will dominate. However, in the region of parameter space where  the masses are tuned so that $(2m_\phi + m_A ) \sim m_S$, the $3 \to 2$ assisted annihilation, now set at resonance, can dominantly drive freeze-out to saturate the required relic density bound. To illustrate the effect of resonance on relic density we define following parameter \cite{Choi:2017mkk} 
\begin{equation}
\delta \equiv \frac{m_S^2-\left(2 m_\phi + m_A\right)^2 }{\left(2 m_\phi + m_A\right)^2 }
\label{eq:delta}
\end{equation}
The thermally averaged cross section near the pole within the narrow width approximation is given by \cite{Choi:2017mkk},
\begin{equation}
\langle \sigma v^2 \rangle \approx \frac{243 \pi \lambda_3 ^2 \,\delta^2 x^3 e^{-3x \delta/2 } }{64  \left(2m_{\phi} + m_A\right)^2 m^2_S \sqrt{m^2_S-4 m^2_{\phi}}}  \rm Br(S \to AA/ \gamma \gamma)\Theta(\delta),
\end{equation}
where $\mu$ has been assumed to be small to suppress the contribution of $2 \to 2$ processes.
In figure \ref{fig:resonance} the relic density is plotted as a function of $\delta$, keeping $m_{\phi} = 200$ MeV, $\epsilon = 0.5$, $\lambda =10^{-5}$, $\mu/m_{\phi}= 10^{-5}, ~ c_{\gamma}=10^{-11}$  MeV$^{-1}$.  The relic density keeping only the corresponding $2 \to 2$ processes in equation \eqref{eq:boltzy} is shown by the light blue lines. The solid, dashed and dot dashed lines corresponds to $\lambda_3=1,~0.5,~0.1$ respectively. The black solid band shows allowed range  of DM  relic density ($\Omega h^2 = 0.12 \pm 0.001$) \cite{Aghanim:2018eyx}. Clearly, a resonant $3 \to 2$ assisted annihilation can effectively drive freeze-out to obtain the required relic density.

\begin{figure}[t]
\begin{center}
\includegraphics[scale=0.3]{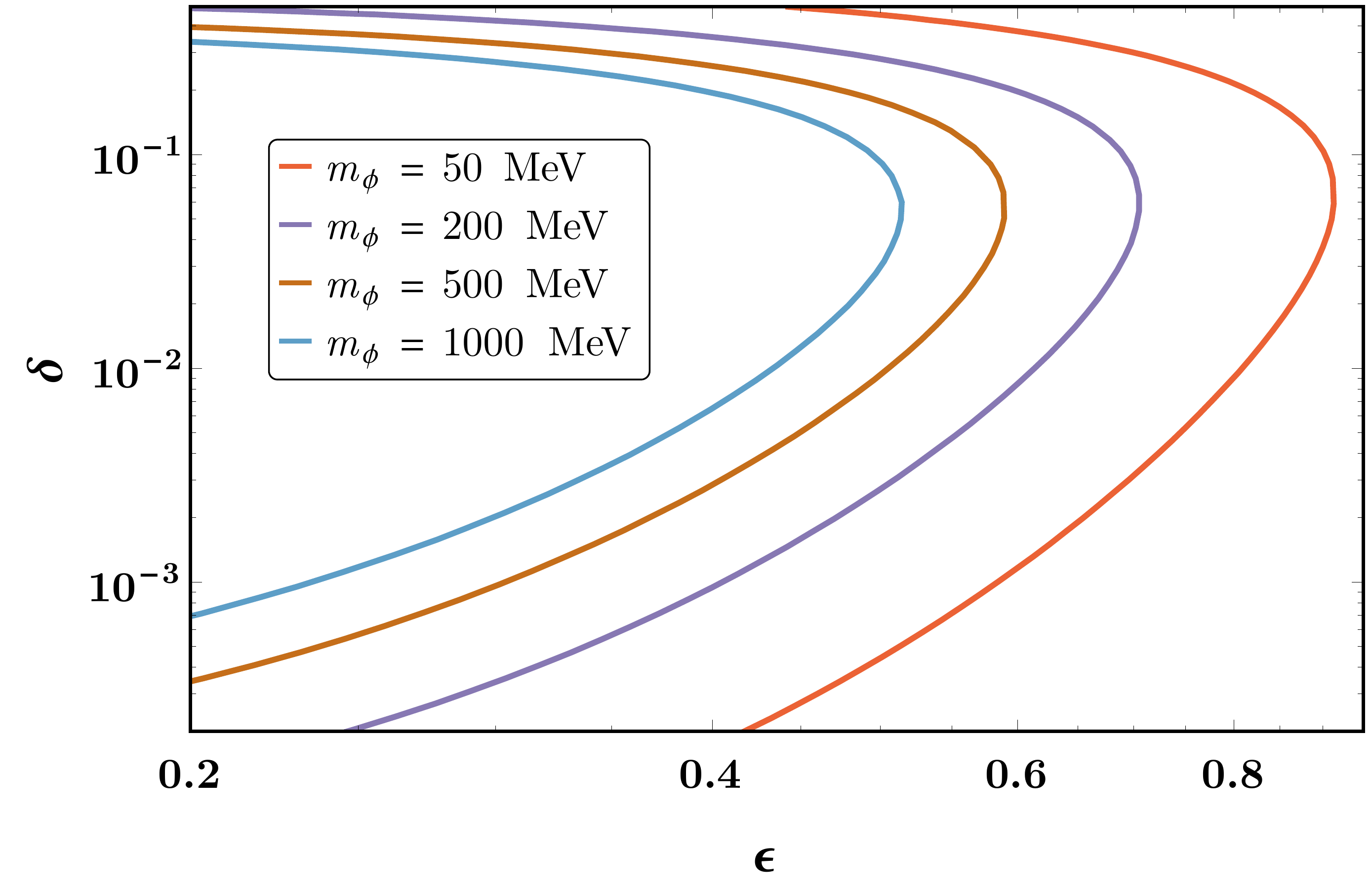}
\caption{Relic density allowed contours in $\epsilon$ vs $\delta $ plane. The light red, light violet, light brown and sky blue lines corresponds to $m_{\phi} = 50$ MeV, $m_{\phi} = 200$ MeV, $m_{\phi} = 500$ MeV and $m_{\phi} = 1000$ MeV respectively. The other couplings are: $\lambda_3 =3, ~\lambda =10^{-5}$, $\mu/m_{\phi}=10^{-5}, ~ c_{\gamma}=10^{-11}$  MeV$^{-1}$.}
\label{fig:relic}
\end{center}
\end{figure}

As is evident from the definition in equation \eqref{eq:delta}, $\delta$ determines how close a parameter point is to resonance and therefore is a measure of tuning in the theory. As $\delta$ is set near the  resonance,  the $3 \to2$  assisted annihilation contribution to the relic density starts dominating while the contribution of the $2 \to 2$ processes become numerically insignificant as evidenced in figure \ref{fig:resonance}. In figure \ref{fig:relic}, DM relic density allowed contours for DM masses $m_{\phi}=50$, $200$, $500$ and $1000$ MeV  has been displayed in $\epsilon - \delta$ plane. In the plot we set $\lambda_3=3$ keeping it safely within the tree level perturbativity limit of $4 \pi$.

\section{Phenomenology}
\label{sec:pheno}
 Being immune to direct detection experiments  the  $3 \to 2$  assisted annihilation  framework is amenable to  probing through indirect effects. First, we examine the  cosmological implications of the light states, especially the late decay of the photophilic MeV scale assisters. Beam dump experiments  can put  complementary constraints on the photophilic assisters. Finally,  the   $\gamma$-ray flux  arising from associated DM annihilation at present day universe may be of interest in the  context of  indirect detection experiments.   We  now elaborate on these  phenomenological consequences of the resonant assisted annihilation framework in the context of the  model presented in section \ref{sec:model}. 

\subsection{Cosmological Constraint}
\label{subsec:bbn}
Any coupling between the MeV scale assister/mediator  $(A/S)$  to the photon will have several cosmological implications. A detailed discussion on this can be found in \cite{Dey:2018yjt}. The late time decay of the $A/S$ to photons may lead to photo-dissociation of the  BBN products \cite{Protheroe:1994dt, Kawasaki:1994sc,Cyburt:2002uv, Jedamzik:2006xz,Poulin:2015opa, Hufnagel:2018bjp, Forestell:2018txr}. This essentially puts an upper bound on the lifetime of the decaying species. Here we use a conservative limit on the lifetime of both the assister and the mediator, to be less than $1$ s, and fix $c_{\gamma}$ to $10^{-11}$  MeV$^{-1}$ which is also consistent with beam dump experiments discussed next.  Additionally,  light degrees of freedom can increase Hubble expansion rate which may alter BBN yields, constraining  the masses  to be greater than $1$ MeV \cite{Cyburt:2015mya}. Other than these, the direct annihilation to photons after neutrino decoupling may change photon to neutrino temperature ratio \cite{Kolb:1986nf, Serpico:2004nm, Nollett:2013pwa, Nollett:2014lwa, Depta:2019lbe}. Following \cite{Depta:2019lbe} we find  that with two light real scalar states in the dark sector, a DM mass below $\sim 8$ MeV is constrained from BBN and CMB observations. 

In the resonant assisted annihilation dominated regime the associated $2 \to 2$ annihilation can inject energy during dark ages which could modify the anisotropies of CMB through ionizing particles. The limit can be presented through the following parameter 
\begin{equation}
p_{\rm ann}= f_{\rm eff} \frac{\langle \sigma v\rangle}{m_{\phi}},
\label{eq:pann}
\end{equation}  
which determines the amount of energy deposited through DM annihilations. The weighted efficiency factor ($f_{\rm eff}$) has been calculated using only the photon spectra \cite{Slatyer:2015jla},  with a conservative choice $f^{\gamma}_{\rm eff} (E)=1$. The most robust bound on $p_{\rm ann}<3.2 \times 10^{-28} \rm cm^3 \rm s^{-1} \rm GeV^{-1}$ has been given by the Planck result \cite{Aghanim:2018eyx}. For $\epsilon=0.5$ the light blue shaded region in figure \ref{fig:lamda-m} has been excluded by aforementioned bound.
\subsection{Fixed Target Searches}
\label{subsec:fixtarget}

An alternate strategy to search for photophilic $A/S$ is through fixed target experiments. There are several experiments like P\textsc{rim}E\textsc{x} \cite{Aloni:2019ruo}, G\textsc{lue}X \cite{Aloni:2019ruo}, E137 \cite{Bjorken:1988as}, Belle-II \cite{Dolan:2017osp}, SHiP \cite{Alekhin:2015byh}, FASER2 \cite{Feng:2018noy}, SeaQuest \cite{Berlin:2018pwi}, and NA 62  \cite{Dobrich:2015jyk} which may probe the effective photon  coupling $c_{\gamma}$  of $A/S.$  A conservative limit of  $c_{\gamma} \leq 10^{-11}$ MeV$^{-1}$ is found to be in consonance with the exclusion bounds in the mass range of interest \cite{Dey:2018yjt}.

\subsection{Indirect Detection}
\label{subsec:ID}
In our region of interest the dominant contribution to the relic density is driven by the $s$-channel resonant assisted annihilation processes, shown in figure \ref{sf:2phiAtoAA} and \ref{sf:2phiAto2g}. However, this $3 \to 2$  annihilation processes is inoperative once the assister number density plummets due to decay. Interestingly, the $2 \to 2$ sub-dominant processes given in figure \ref{sf:2phito2A} and \ref{sf:2phito2g} survives and provides a handle to explore these framework through indirect detection. In the most generic form of the Lagrangian in equation \eqref{eq:pot1} it is possible to tune the couplings to drive resonant assisted annihilation while keeping all the relevant cross sections negligible. However, in the benchmark model due to universal coupling choices  a sizeable  assisted annihilation would lead to a correlated cross section of indirect detection. It is in this context we explore the possibility to probe the parameter space of the benchmark model through indirect detection.

For the presented model, photophilic assister may produce potential $\gamma$-ray flux. The differential photon flux from the annihilation of a self-conjugate DM is given by \cite{Slatyer:2017sev}
\begin{equation}
\label{eq:ID-flux1}
\Phi_{\gamma}^{\prime} \left(E_{\gamma} \right) = \frac{\rho_{\odot}^2 r_{\odot}}{8 \pi m_{\phi}^2} \sum_{i} \langle \sigma v \rangle_{i} \,  \frac{d N^{i}_{\gamma}}{dE_{\gamma}} \frac{J}{\Delta\Omega},
\end{equation} 
\begin{figure}[t]
\begin{center}
\textbf{$\boldsymbol{\epsilon < 1}$ \hspace{6.5 cm}	$\boldsymbol{\epsilon > 1}$}
\subfloat[\label{sf:Eg-phig-lt} Convoluted gamma-ray flux $\Phi_{\gamma} \left(E_{\gamma} \right)$  for two DM masses of $10$ MeV and $500$ MeV by light blue and light green lines respectively. The solid lines represent $\epsilon = 0.99$ while dashed lines are for $\epsilon = 0.5$.]{
\includegraphics[scale=0.25]{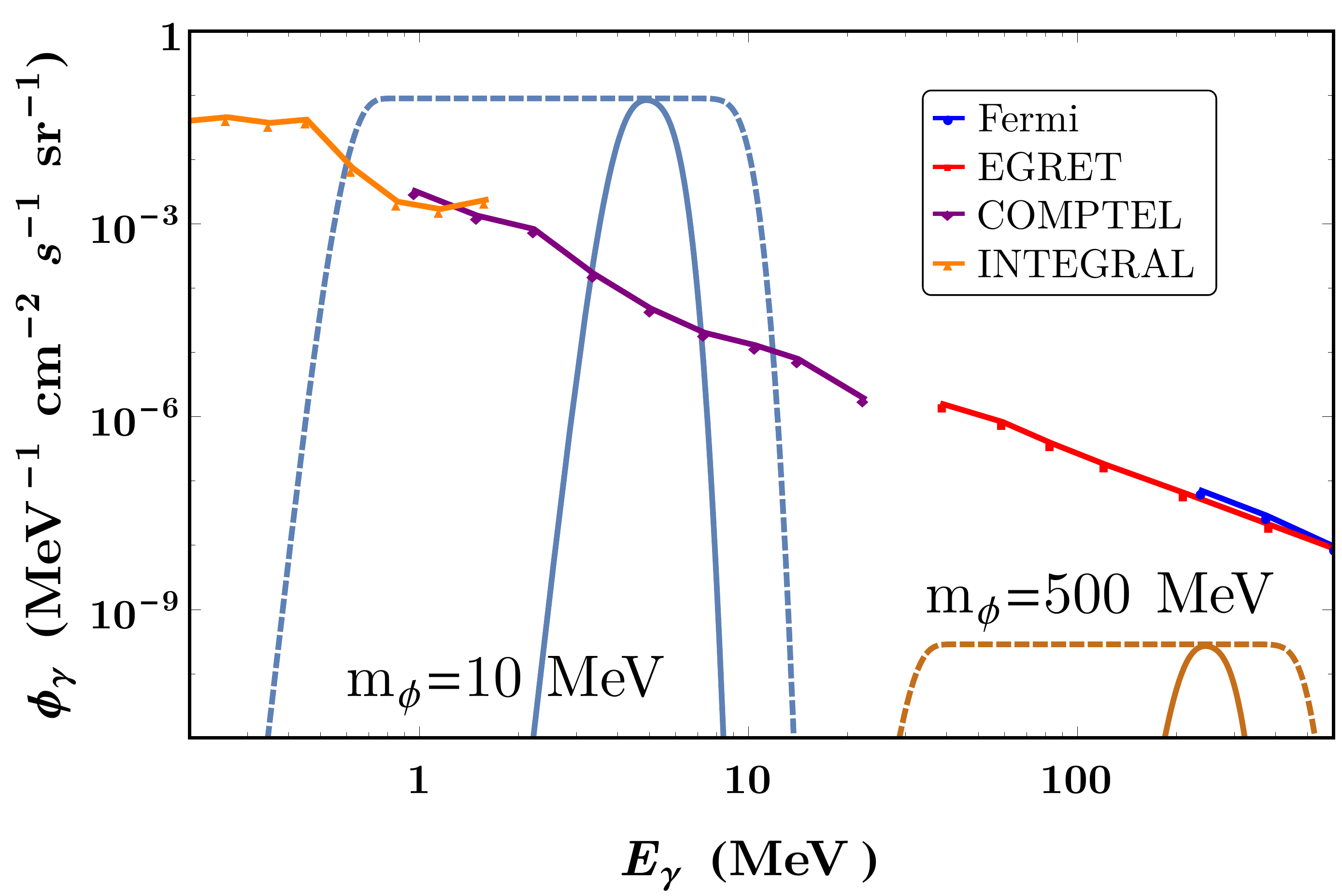}}~~~~
\subfloat[\label{sf:Eg-phig-gt} Smeared gamma-ray flux $\Phi_{\gamma} \left(E_{\gamma} \right)$ for two DM masses of $10$ MeV and $200$ MeV with $\epsilon >1$. ]{
\includegraphics[scale=0.26]{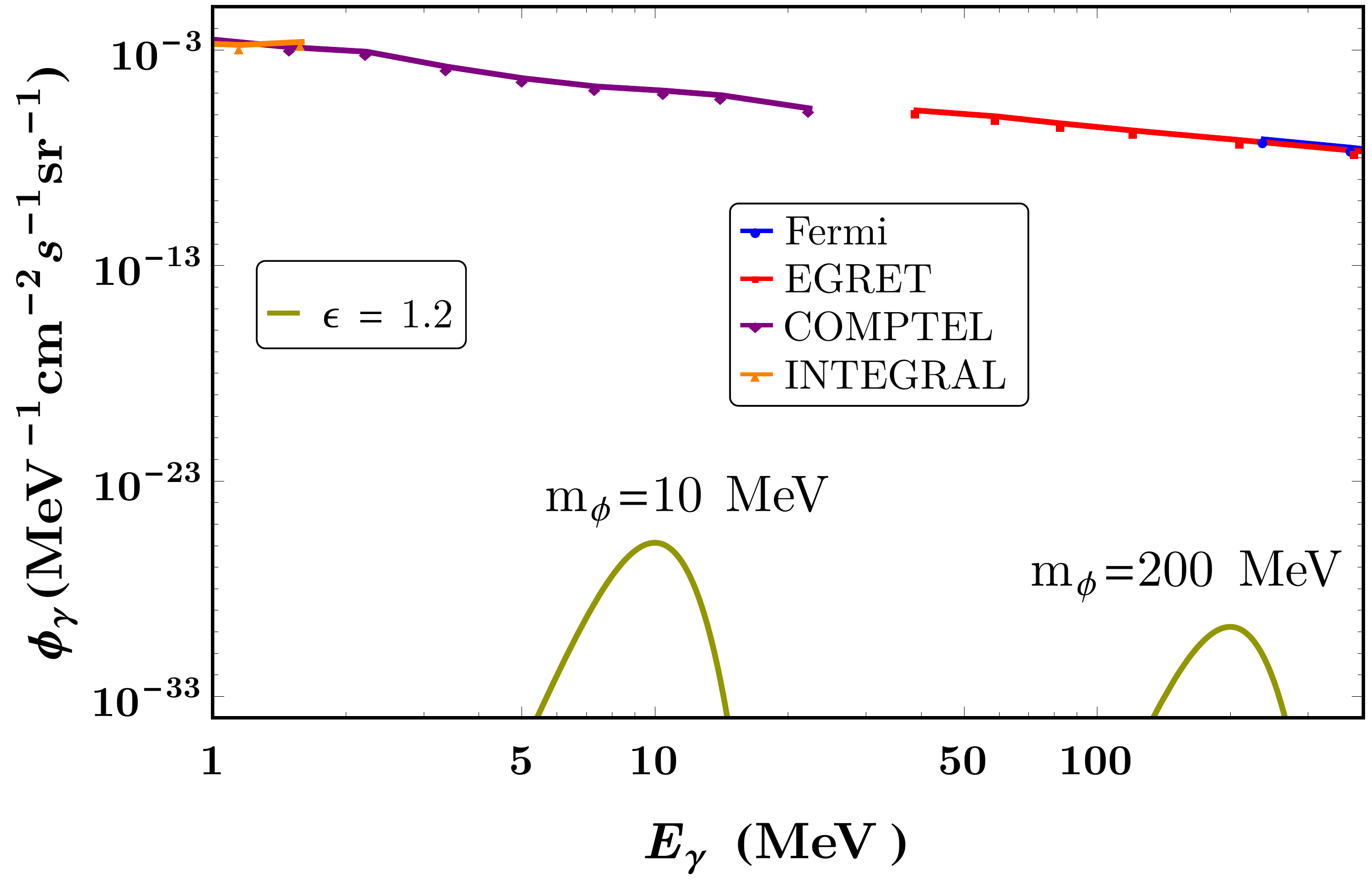}}
\caption{The chosen values of the remaining parameters are mentioned in figure \ref{fig:relic}.}
\label{fig:ID-flux}
\end{center}
\end{figure} 

where $\langle \sigma v \rangle_i$ is the thermally averaged annihilation cross section, $d N^i_{\gamma}/dE_{\gamma}$ is the corresponding spectrum of $\gamma$-rays, $r_{\odot} \simeq 8.5$ kpc is the Sun's distance from the Galactic center, $\rho_{\odot} \simeq 0.3$ GeV/cm$^3$ is the local DM density, and $J$ is the standard $J$-factor which integrates intermediate DM density along the line of sight over the solid angel $\Delta\Omega$. We have used NFW profile \cite{Navarro:1995iw, Navarro:1996gj} to calculate $J$ factor for the considered indirect detection experiments \cite{Essig:2013goa}. The dominant contribution on the $\gamma$-ray flux comes from two different kinds of processes
\begin{enumerate}
\item Two body annihilation to assisters as shown in figure \ref{sf:2phito2A}. In the center-of-mass frame of the DM, the subsequent decay of the assisters to photons will develop a box-shaped spectra for the former. The spectra of the photon can be written as  \cite{Ibarra:2012dw, Boddy:2015efa} 
\begin{equation}
\frac{d N_{\gamma}}{dE_{\gamma}} = \frac{4}{\Delta E} \, \Theta \left(E_{\gamma} - E_{-}\right) \Theta \left(E_{\gamma} - E_{+}\right),
\label{eq:box}
\end{equation}
where 
\begin{equation*}
E_{\pm}=\frac{m_{\phi}}{2} \left(1\pm\sqrt{1-\frac{m_A^2}{m^2_{\phi}}} \right)
\end{equation*}
are the edges of the box and $\Delta E$ is the difference between them and $\Theta$ is the step function. This channel will be operative only for $\epsilon < 1$.

\item Direct two body annihilation to photons as depicted in figure \ref{sf:2phito2g}, which will be functional both for $\epsilon < 1$ and $\epsilon > 1$. In the center-of-mass frame of the DM, this would give rise to following line spectra of the photons
\begin{equation}
\frac{d N_{\gamma}}{dE_{\gamma}} = 2 \, {\mathcal{\delta}}\left(E_{\gamma}-m_{\phi} \right),
\label{eq:line}
\end{equation}
\end{enumerate}

where $\delta\left(E_{\gamma}-m_{\phi} \right)$ is the Dirac delta function.  We have assumed a Gaussian detector response \cite{Bringmann:2008kj} which would spread out the spikes and sharp  kinematic edges of the flux ($\Phi_{\gamma}^{\prime} \left(E_{\gamma} \right)$) and the the convoluted gamma-ray flux ($\Phi_{\gamma} \left(E_{\gamma} \right)$) has been compared with the experimental results.  There have been several gamma-ray satellites which search for such flux from DM annihilation. Since we are exploring  phenomenology of DM mass in sub-GeV range, therefore  we have used outcome  of low energy gamma-ray detector like HEAO-1 \cite{Gruber:1999yr}, INTEGRAL \cite{Bouchet:2008rp}, COMPTEL \cite{Weidenspointner:99, Kappadath:98}, EGRET \cite{Strong:2004de} and Fermi \cite{Ackermann:2012pya} to obtain the constraint on the relevant parameters of the model presented in section \ref{sec:model}. We have used central values of the observations of the experiments and interpolated that to obtain a continuous flux in their respective energy window.
\begin{figure}[t]
\begin{center}
\includegraphics[scale=0.3]{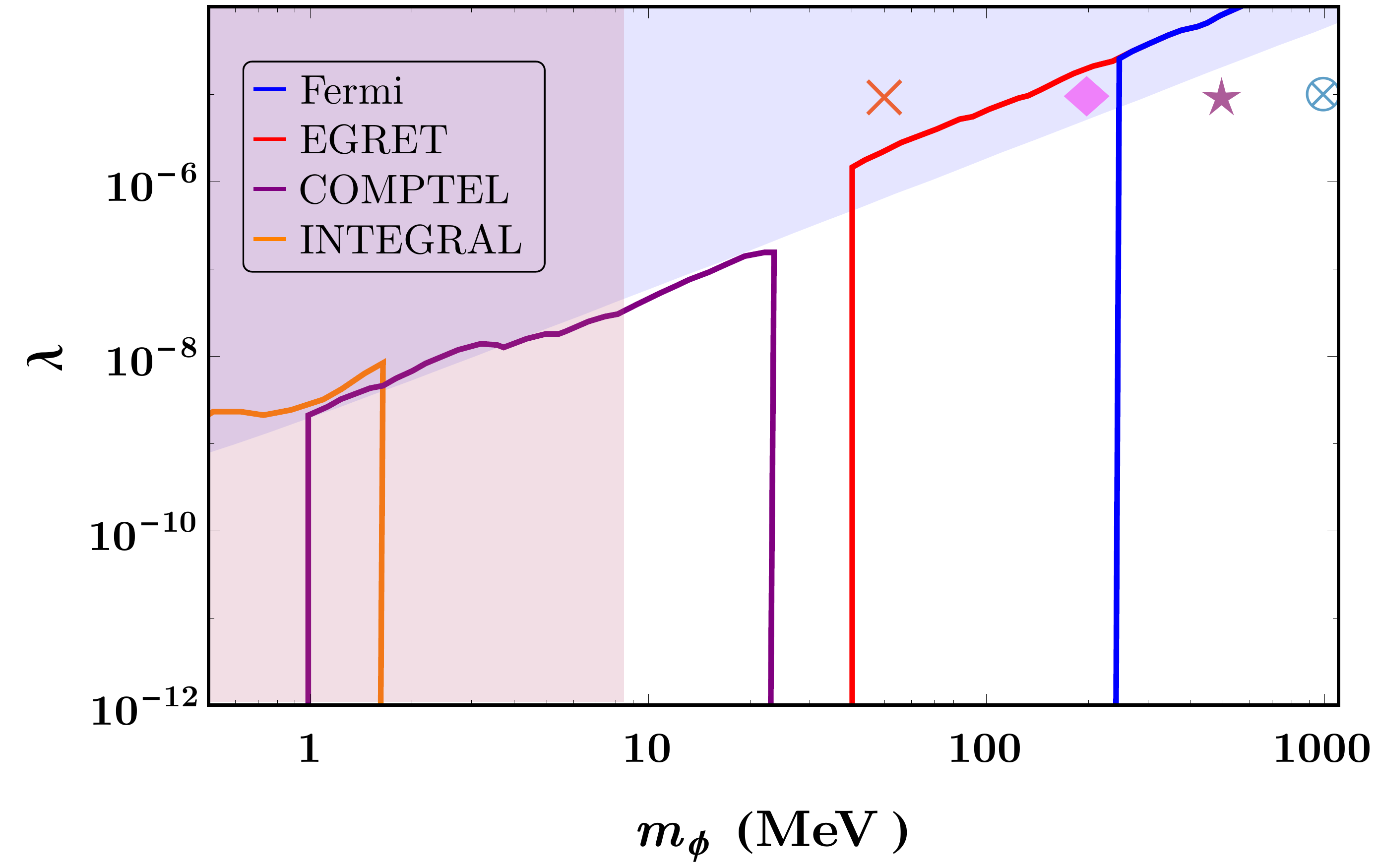}
\caption{Allowed regions of $\lambda$ as a function of DM mass $m_{\phi}$. The upper limit on $\lambda$ for $\epsilon=0.5$ from INTEGRAL, COMPTEL, EGRET and Fermi is shown by orange, purple, red, and blue lines respectively. The BBN and CMB excluded region is shown by light red shading. The light blue shaded region denotes CMB exclusion region from the energy injection through DM annihilation. The other parameters are the same as in figure \ref{fig:relic}.}
\label{fig:lamda-m}
\end{center}
\end{figure}    

In figures \ref{sf:Eg-phig-lt} and \ref{sf:Eg-phig-gt} we have shown smeared differential gamma-ray flux ($\Phi_{\gamma} \left(E_{\gamma} \right)$) as a function of gamma-ray energy for $\epsilon <1$ and $\epsilon >1$ respectively. For $\epsilon <1$ both $\phi \phi \to 4 \gamma$ through $A$ and $\phi \phi \to \gamma \gamma$ is operative. However, from BBN and fixed target constraints on $c_{\gamma}$ and for $\mu/m_{\phi} = 10^{-5}$ contribution of former to the total flux is numerically insignificant, this leads to box type spectrum as shown in figure \ref{sf:Eg-phig-lt}. Apart from this, with the choice $\lambda = \mu/m_{\phi}=10^{-5}$, DM annihilation to assisters through point interaction would dominate the same annihilation through $A/S$ mediation. For $\epsilon > 1$ only channel to probe DM signals through indirect detection is $\phi \phi \to \gamma \gamma$.  However, in the region of parameter space where resonance assisted annihilation is the dominant channel to drive the freeze-out,  smallness of $\mu$ and $c_{\gamma}$ makes the differential gamma-ray flux beyond the reach of the current experimental sensitivity. This has been shown in figure \ref{sf:Eg-phig-gt} for DM masses $10$ and $200$ MeV. Since the thermally averaged cross section $\langle \sigma v \rangle_i$ of a particular channel is inversely proportional to $m_{\phi}^2$, therefore for a fixed choice of  other couplings and masses the maximum value of flux increases with decrease in DM mass.

The upper limit on $\lambda$ for the benchmark model from INTEGRAL, COMPTEL, EGRET and Fermi against the DM mass is shown in figure \ref{fig:lamda-m}. For completeness, four benchmark points which satisfy the relic density constraint are shown by circle with cross, star, diamond, and cross. The details of the benchmark points are given in  table  \ref{tab:benchmark}.

\begin{table}[t]\caption{Benchmark points}
\label{tab:benchmark}
\centering  
\begin{tabular}{|c|c|c|c|c|c|c|c|c|}
\hline\hline  
BP           &  $m_{\phi}$     & $\lambda$ & $\lambda_3$ & $c_{\gamma}$ & $\mu/ m_{\phi}$   & $\epsilon$  &  $\delta$      & Flux $ \Phi_{\gamma} \left(E_{+}  \right)$ \\ 
             & (MeV)           &     &           &  MeV$^{-1}$   &     &         &          
& (MeV s sr)$^{-1}$  cm$^{-2}$   \\
\hline
\multirow{2}*{BP1 \textbf{$\times$}}     & \multirow{2}*{50 }   &
\multirow{2}*{$10^{-5}$} & \multirow{2}*{$3$} & \multirow{2}*{$10^{-11}$}  
 & \multirow{2}*{$10^{-5}$}  &\multirow{2}*{ $0.5$ }      & $0.445 $ & \multirow{2}*{$1.3 \times 10^{-5}$ }\\
\cline{8-8}   
 &   &   &  &  &    &  & $4.47 \times 10^{-4}$ &  \\

\hline

\multirow{2}*{BP2 $\blacklozenge$}     & \multirow{2}*{200 }  
& \multirow{2}*{$10^{-5}$}  & \multirow{2}*{$3$} & \multirow{2}*{$10^{-11}$} 
 & \multirow{2}*{$10^{-5}$}   &  \multirow{2}*{$0.5$}     &  $0.282$   & \multirow{2}*{$1.3 \times 10^{-8}$}  \\  
\cline{8-8}
 &     &   &  &   &    &      &  $2.79 \times 10^{-3}$ & \\

\hline

\multirow{2}*{BP3 $\bigstar$}     & \multirow{2}*{500 }   
& \multirow{2}*{$10^{-5}$}  &\multirow{2}*{$3$} &  \multirow{2}*{$10^{-11}$} 
& \multirow{2}*{$10^{-5}$} & \multirow{2}*{$0.5$} & $0.174$ &\multirow{2}*{ $1.3 \times 10^{-10}$}  \\ 
 \cline{8-8}
 &   &   &  &  &    &  & $9.60 \times 10^{-3}$ & \\

\hline

\multirow{2}*{BP4 $\bigotimes$}     & \multirow{2}*{1000 }   
& \multirow{2}*{$10^{-5}$}  & \multirow{2}*{$3$} & \multirow{2}*{$10^{-11}$} 
& \multirow{2}*{$10^{-5}$} & \multirow{2}*{$0.5$} & $0.088$ &\multirow{2}*{ $4.2 \times 10^{-12}$}  \\ 
 \cline{8-8}
 &   &  &  &    &    &  & $0.030$ &  \\
 \hline \hline
\end{tabular}
\end{table}

\section{Conclusions}
\label{sec:summary}
In this paper, we present an alternate possibility to drive thermal freeze-out, through a multi-body $3 \to 2$ resonant assisted annihilation. Here along with a pair of DM,  there is an SM-like assister in the initial state. The key challenge here is to overcome the contribution of related $2 \to 2$ channels in comparison to these flux suppressed multi-body $3 \to 2$ assisted annihilation processes. In this paper, we have presented a simple  model having three real  scalars, a  stable DM ($\phi$), an assister ($A$) and a heavy mediator ($S$) where the latter two also double up as a portal to the visible sector. We find that in the region of parameter space where masses are tuned to $(2m_\phi + m_A ) \sim m_S$ an $s-$channel $3 \to 2$ assisted annihilation channel can have resonance and dominantly drive freeze-out to produce the observed relic density of DM. We show that in this tuned region the relic abundance is in the right ballpark for a DM mass between  few MeV to a few GeV with perturbative couplings.  

The resonant assisted annihilation channels are difficult to probe in direct detection experiments owing to its novel topology. In this article, we have shown that even in the distinctive resonance region, the correlated $2 \to 2$ annihilation channels can produce appreciable indirect detection signatures.  Annihilation of the DM to assisters and subsequent decay of the photophilic assister can be constrained  by experiments like INTEGRAL, EGRET, COMPTEL, Fermi, and through the anisotropies of CMB   in certain regions of the parameter space.

\paragraph*{Acknowledgments\,:} 
We would like to thank Ujjal Kumar Dey for comments on the manuscript. TNM would like to thank MHRD, Government of India for a research fellowship. TSR is partially supported by the Department of Science and Technology, Government of India, under the Grant Agreement No. IFA13-PH-74 (INSPIRE Faculty Award).

%

\bibliographystyle{JHEP}
\bibliography{resonance-ref.bib}

\providecommand{\href}[2]{#2}\begingroup\raggedright\begin{thebibliography}{10}

\bibitem{Spergel:1999mh}
D.~N. Spergel and P.~J. Steinhardt, \emph{{Observational evidence for
  selfinteracting cold dark matter}},
  \href{http://dx.doi.org/10.1103/PhysRevLett.84.3760}{\emph{Phys. Rev. Lett.}
  {\bf 84} (2000) 3760--3763},
  [\href{https://arxiv.org/abs/astro-ph/9909386}{{\tt astro-ph/9909386}}].

\bibitem{Nakama:2017ohe}
T.~Nakama, J.~Chluba and M.~Kamionkowski, \emph{{Shedding light on the
  small-scale crisis with CMB spectral distortions}},
  \href{http://dx.doi.org/10.1103/PhysRevD.95.121302}{\emph{Phys. Rev.} {\bf
  D95} (2017) 121302}, [\href{https://arxiv.org/abs/1703.10559}{{\tt
  1703.10559}}].

\bibitem{Bullock:2017xww}
J.~S. Bullock and M.~Boylan-Kolchin, \emph{{Small-Scale Challenges to the
  $\Lambda$CDM Paradigm}},
  \href{http://dx.doi.org/10.1146/annurev-astro-091916-055313}{\emph{Ann. Rev.
  Astron. Astrophys.} {\bf 55} (2017) 343--387},
  [\href{https://arxiv.org/abs/1707.04256}{{\tt 1707.04256}}].

\bibitem{Crisler:2018gci}
{\scshape SENSEI} collaboration, M.~Crisler, R.~Essig, J.~Estrada,
  G.~Fernandez, J.~Tiffenberg, M.~Sofo~haro et~al., \emph{{SENSEI: First
  Direct-Detection Constraints on sub-GeV Dark Matter from a Surface Run}},
  \href{http://dx.doi.org/10.1103/PhysRevLett.121.061803}{\emph{Phys. Rev.
  Lett.} {\bf 121} (2018) 061803},
  [\href{https://arxiv.org/abs/1804.00088}{{\tt 1804.00088}}].

\bibitem{Abramoff:2019dfb}
{\scshape SENSEI} collaboration, O.~Abramoff et~al., \emph{{SENSEI:
  Direct-Detection Constraints on Sub-GeV Dark Matter from a Shallow
  Underground Run Using a Prototype Skipper-CCD}},
  \href{http://dx.doi.org/10.1103/PhysRevLett.122.161801}{\emph{Phys. Rev.
  Lett.} {\bf 122} (2019) 161801},
  [\href{https://arxiv.org/abs/1901.10478}{{\tt 1901.10478}}].

\bibitem{Essig:2011nj}
R.~Essig, J.~Mardon and T.~Volansky, \emph{{Direct Detection of Sub-GeV Dark
  Matter}}, \href{http://dx.doi.org/10.1103/PhysRevD.85.076007}{\emph{Phys.
  Rev.} {\bf D85} (2012) 076007}, [\href{https://arxiv.org/abs/1108.5383}{{\tt
  1108.5383}}].

\bibitem{Essig:2015cda}
R.~Essig, M.~Fernandez-Serra, J.~Mardon, A.~Soto, T.~Volansky and T.-T. Yu,
  \emph{{Direct Detection of sub-GeV Dark Matter with Semiconductor Targets}},
  \href{http://dx.doi.org/10.1007/JHEP05(2016)046}{\emph{JHEP} {\bf 05} (2016)
  046}, [\href{https://arxiv.org/abs/1509.01598}{{\tt 1509.01598}}].

\bibitem{Lee:2015qva}
S.~K. Lee, M.~Lisanti, S.~Mishra-Sharma and B.~R. Safdi, \emph{{Modulation
  Effects in Dark Matter-Electron Scattering Experiments}},
  \href{http://dx.doi.org/10.1103/PhysRevD.92.083517}{\emph{Phys. Rev.} {\bf
  D92} (2015) 083517}, [\href{https://arxiv.org/abs/1508.07361}{{\tt
  1508.07361}}].

\bibitem{Hochberg:2016sqx}
Y.~Hochberg, T.~Lin and K.~M. Zurek, \emph{{Absorption of light dark matter in
  semiconductors}},
  \href{http://dx.doi.org/10.1103/PhysRevD.95.023013}{\emph{Phys. Rev.} {\bf
  D95} (2017) 023013}, [\href{https://arxiv.org/abs/1608.01994}{{\tt
  1608.01994}}].

\bibitem{Kurinsky:2019pgb}
N.~A. Kurinsky, T.~C. Yu, Y.~Hochberg and B.~Cabrera, \emph{{Diamond Detectors
  for Direct Detection of Sub-GeV Dark Matter}},
  \href{https://arxiv.org/abs/1901.07569}{{\tt 1901.07569}}.

\bibitem{Hochberg:2015pha}
Y.~Hochberg, Y.~Zhao and K.~M. Zurek, \emph{{Superconducting Detectors for
  Superlight Dark Matter}},
  \href{http://dx.doi.org/10.1103/PhysRevLett.116.011301}{\emph{Phys. Rev.
  Lett.} {\bf 116} (2016) 011301},
  [\href{https://arxiv.org/abs/1504.07237}{{\tt 1504.07237}}].

\bibitem{Hochberg:2016ajh}
Y.~Hochberg, T.~Lin and K.~M. Zurek, \emph{{Detecting Ultralight Bosonic Dark
  Matter via Absorption in Superconductors}},
  \href{http://dx.doi.org/10.1103/PhysRevD.94.015019}{\emph{Phys. Rev.} {\bf
  D94} (2016) 015019}, [\href{https://arxiv.org/abs/1604.06800}{{\tt
  1604.06800}}].

\bibitem{Hochberg:2019cyy}
Y.~Hochberg, I.~Charaev, S.-W. Nam, V.~Verma, M.~Colangelo and K.~K. Berggren,
  \emph{{Detecting Dark Matter with Superconducting Nanowires}},
  \href{https://arxiv.org/abs/1903.05101}{{\tt 1903.05101}}.

\bibitem{Dror:2019onn}
J.~A. Dror, G.~Elor and R.~Mcgehee, \emph{{Direct Detection Signals from
  Absorption of Fermionic Dark Matter}},
  \href{https://arxiv.org/abs/1905.12635}{{\tt 1905.12635}}.

\bibitem{Battaglieri:2017aum}
M.~Battaglieri et~al., \emph{{US Cosmic Visions: New Ideas in Dark Matter 2017:
  Community Report}},  in \emph{{U.S. Cosmic Visions: New Ideas in Dark Matter
  College Park, MD, USA, March 23-25, 2017}}, 2017.
\newblock \href{https://arxiv.org/abs/1707.04591}{{\tt 1707.04591}}.

\bibitem{Dolgov:1980uu}
A.~D. Dolgov, \emph{{ON CONCENTRATION OF RELICT THETA PARTICLES. (IN
  RUSSIAN)}}, {\emph{Yad. Fiz.} {\bf 31} (1980) 1522--1528}.

\bibitem{Dolgov:2017ujf}
A.~D. Dolgov, \emph{{New Old Mechanism of Dark Matter Burning}},
  \href{https://arxiv.org/abs/1705.03689}{{\tt 1705.03689}}.

\bibitem{Hochberg:2014dra}
Y.~Hochberg, E.~Kuflik, T.~Volansky and J.~G. Wacker, \emph{{Mechanism for
  Thermal Relic Dark Matter of Strongly Interacting Massive Particles}},
  \href{http://dx.doi.org/10.1103/PhysRevLett.113.171301}{\emph{Phys. Rev.
  Lett.} {\bf 113} (2014) 171301}, [\href{https://arxiv.org/abs/1402.5143}{{\tt
  1402.5143}}].

\bibitem{Hochberg:2014kqa}
Y.~Hochberg, E.~Kuflik, H.~Murayama, T.~Volansky and J.~G. Wacker, \emph{{Model
  for Thermal Relic Dark Matter of Strongly Interacting Massive Particles}},
  \href{http://dx.doi.org/10.1103/PhysRevLett.115.021301}{\emph{Phys. Rev.
  Lett.} {\bf 115} (2015) 021301}, [\href{https://arxiv.org/abs/1411.3727}{{\tt
  1411.3727}}].

\bibitem{Dey:2016qgf}
U.~K. Dey, T.~N. Maity and T.~S. Ray, \emph{{Light Dark Matter through Assisted
  Annihilation}},
  \href{http://dx.doi.org/10.1088/1475-7516/2017/03/045}{\emph{JCAP} {\bf 1703}
  (2017) 045}, [\href{https://arxiv.org/abs/1612.09074}{{\tt 1612.09074}}].

\bibitem{Bernal:2015xba}
N.~Bernal and X.~Chu, \emph{{$\mathbb {Z}_2$ SIMP Dark Matter}},
  \href{http://dx.doi.org/10.1088/1475-7516/2016/01/006}{\emph{JCAP} {\bf 1601}
  (2016) 006}, [\href{https://arxiv.org/abs/1510.08527}{{\tt 1510.08527}}].

\bibitem{Lee:2015uva}
H.~M. Lee and M.-S. Seo, \emph{{Models for SIMP dark matter and dark photon}},
  \href{http://dx.doi.org/10.1063/1.4953320}{\emph{AIP Conf. Proc.} {\bf 1743}
  (2016) 060003}, [\href{https://arxiv.org/abs/1510.05116}{{\tt 1510.05116}}].

\bibitem{Choi:2015bya}
S.-M. Choi and H.~M. Lee, \emph{{SIMP dark matter with gauged Z$_{3}$
  symmetry}}, \href{http://dx.doi.org/10.1007/JHEP09(2015)063}{\emph{JHEP} {\bf
  09} (2015) 063}, [\href{https://arxiv.org/abs/1505.00960}{{\tt 1505.00960}}].

\bibitem{Bernal:2015bla}
N.~Bernal, C.~Garcia-Cely and R.~Rosenfeld, \emph{{WIMP and SIMP Dark Matter
  from the Spontaneous Breaking of a Global Group}},
  \href{http://dx.doi.org/10.1088/1475-7516/2015/04/012}{\emph{JCAP} {\bf 1504}
  (2015) 012}, [\href{https://arxiv.org/abs/1501.01973}{{\tt 1501.01973}}].

\bibitem{Hochberg:2015vrg}
Y.~Hochberg, E.~Kuflik and H.~Murayama, \emph{{SIMP Spectroscopy}},
  \href{http://dx.doi.org/10.1007/JHEP05(2016)090}{\emph{JHEP} {\bf 05} (2016)
  090}, [\href{https://arxiv.org/abs/1512.07917}{{\tt 1512.07917}}].

\bibitem{Kuflik:2015isi}
E.~Kuflik, M.~Perelstein, N.~R.-L. Lorier and Y.-D. Tsai, \emph{{Elastically
  Decoupling Dark Matter}},
  \href{http://dx.doi.org/10.1103/PhysRevLett.116.221302}{\emph{Phys. Rev.
  Lett.} {\bf 116} (2016) 221302},
  [\href{https://arxiv.org/abs/1512.04545}{{\tt 1512.04545}}].

\bibitem{Choi:2016tkj}
S.-M. Choi, Y.-J. Kang and H.~M. Lee, \emph{{On thermal production of
  self-interacting dark matter}},
  \href{http://dx.doi.org/10.1007/JHEP12(2016)099}{\emph{JHEP} {\bf 12} (2016)
  099}, [\href{https://arxiv.org/abs/1610.04748}{{\tt 1610.04748}}].

\bibitem{Choi:2016hid}
S.-M. Choi and H.~M. Lee, \emph{{Resonant SIMP dark matter}},
  \href{http://dx.doi.org/10.1016/j.physletb.2016.04.055}{\emph{Phys. Lett.}
  {\bf B758} (2016) 47--53}, [\href{https://arxiv.org/abs/1601.03566}{{\tt
  1601.03566}}].

\bibitem{Bernal:2017mqb}
N.~Bernal, X.~Chu and J.~Pradler, \emph{{Simply split strongly interacting
  massive particles}},
  \href{http://dx.doi.org/10.1103/PhysRevD.95.115023}{\emph{Phys. Rev.} {\bf
  D95} (2017) 115023}, [\href{https://arxiv.org/abs/1702.04906}{{\tt
  1702.04906}}].

\bibitem{Ho:2017fte}
S.-Y. Ho, T.~Toma and K.~Tsumura, \emph{{A Radiative Neutrino Mass Model with
  SIMP Dark Matter}},
  \href{http://dx.doi.org/10.1007/JHEP07(2017)101}{\emph{JHEP} {\bf 07} (2017)
  101}, [\href{https://arxiv.org/abs/1705.00592}{{\tt 1705.00592}}].

\bibitem{Cline:2017tka}
J.~M. Cline, H.~Liu, T.~Slatyer and W.~Xue, \emph{{Enabling Forbidden Dark
  Matter}}, \href{http://dx.doi.org/10.1103/PhysRevD.96.083521}{\emph{Phys.
  Rev.} {\bf D96} (2017) 083521}, [\href{https://arxiv.org/abs/1702.07716}{{\tt
  1702.07716}}].

\bibitem{Choi:2017zww}
S.-M. Choi, Y.~Hochberg, E.~Kuflik, H.~M. Lee, Y.~Mambrini, H.~Murayama et~al.,
  \emph{{Vector SIMP dark matter}},
  \href{http://dx.doi.org/10.1007/JHEP10(2017)162}{\emph{JHEP} {\bf 10} (2017)
  162}, [\href{https://arxiv.org/abs/1707.01434}{{\tt 1707.01434}}].

\bibitem{Kuflik:2017iqs}
E.~Kuflik, M.~Perelstein, N.~R.-L. Lorier and Y.-D. Tsai, \emph{{Phenomenology
  of ELDER Dark Matter}},
  \href{http://dx.doi.org/10.1007/JHEP08(2017)078}{\emph{JHEP} {\bf 08} (2017)
  078}, [\href{https://arxiv.org/abs/1706.05381}{{\tt 1706.05381}}].

\bibitem{Hochberg:2018rjs}
Y.~Hochberg, E.~Kuflik, R.~Mcgehee, H.~Murayama and K.~Schutz, \emph{{Strongly
  interacting massive particles through the axion portal}},
  \href{http://dx.doi.org/10.1103/PhysRevD.98.115031}{\emph{Phys. Rev.} {\bf
  D98} (2018) 115031}, [\href{https://arxiv.org/abs/1806.10139}{{\tt
  1806.10139}}].

\bibitem{Hochberg:2018vdo}
Y.~Hochberg, E.~Kuflik and H.~Murayama, \emph{{Twin Higgs model with strongly
  interacting massive particle dark matter}},
  \href{http://dx.doi.org/10.1103/PhysRevD.99.015005}{\emph{Phys. Rev.} {\bf
  D99} (2019) 015005}, [\href{https://arxiv.org/abs/1805.09345}{{\tt
  1805.09345}}].

\bibitem{Bhattacharya:2019mmy}
S.~Bhattacharya, P.~Ghosh and S.~Verma, \emph{{SIMPler realisation of Scalar
  Dark Matter}},  \href{https://arxiv.org/abs/1904.07562}{{\tt 1904.07562}}.

\bibitem{Chauhan:2017eck}
B.~Chauhan, \emph{{Sub-MeV Self Interacting Dark Matter}},
  \href{http://dx.doi.org/10.1103/PhysRevD.97.123017}{\emph{Phys. Rev.} {\bf
  D97} (2018) 123017}, [\href{https://arxiv.org/abs/1711.02970}{{\tt
  1711.02970}}].

\bibitem{Choi:2017mkk}
S.-M. Choi, H.~M. Lee and M.-S. Seo, \emph{{Cosmic abundances of SIMP dark
  matter}}, \href{http://dx.doi.org/10.1007/JHEP04(2017)154}{\emph{JHEP} {\bf
  04} (2017) 154}, [\href{https://arxiv.org/abs/1702.07860}{{\tt 1702.07860}}].

\bibitem{Dey:2018yjt}
U.~K. Dey, T.~N. Maity and T.~S. Ray, \emph{{Boosting Assisted Annihilation for
  a Cosmologically Safe MeV Scale Dark Matter}},
  \href{http://dx.doi.org/10.1103/PhysRevD.99.095025}{\emph{Phys. Rev.} {\bf
  D99} (2019) 095025}, [\href{https://arxiv.org/abs/1812.11418}{{\tt
  1812.11418}}].

\bibitem{Aghanim:2018eyx}
{\scshape Planck} collaboration, N.~Aghanim et~al., \emph{{Planck 2018 results.
  VI. Cosmological parameters}},  \href{https://arxiv.org/abs/1807.06209}{{\tt
  1807.06209}}.

\bibitem{Drees:2015exa}
M.~Drees, F.~Hajkarim and E.~R. Schmitz, \emph{{The Effects of QCD Equation of
  State on the Relic Density of WIMP Dark Matter}},
  \href{http://dx.doi.org/10.1088/1475-7516/2015/06/025}{\emph{JCAP} {\bf 1506}
  (2015) 025}, [\href{https://arxiv.org/abs/1503.03513}{{\tt 1503.03513}}].

\bibitem{Aloni:2019ruo}
D.~Aloni, C.~Fanelli, Y.~Soreq and M.~Williams, \emph{{Photoproduction of
  axion-like particles}},  \href{https://arxiv.org/abs/1903.03586}{{\tt
  1903.03586}}.

\bibitem{Protheroe:1994dt}
R.~J. Protheroe, T.~Stanev and V.~S. Berezinsky, \emph{{Electromagnetic
  cascades and cascade nucleosynthesis in the early universe}},
  \href{http://dx.doi.org/10.1103/PhysRevD.51.4134}{\emph{Phys. Rev.} {\bf D51}
  (1995) 4134--4144}, [\href{https://arxiv.org/abs/astro-ph/9409004}{{\tt
  astro-ph/9409004}}].

\bibitem{Kawasaki:1994sc}
M.~Kawasaki and T.~Moroi, \emph{{Electromagnetic cascade in the early universe
  and its application to the big bang nucleosynthesis}},
  \href{http://dx.doi.org/10.1086/176324}{\emph{Astrophys. J.} {\bf 452} (1995)
  506}, [\href{https://arxiv.org/abs/astro-ph/9412055}{{\tt
  astro-ph/9412055}}].

\bibitem{Cyburt:2002uv}
R.~H. Cyburt, J.~R. Ellis, B.~D. Fields and K.~A. Olive, \emph{{Updated
  nucleosynthesis constraints on unstable relic particles}},
  \href{http://dx.doi.org/10.1103/PhysRevD.67.103521}{\emph{Phys. Rev.} {\bf
  D67} (2003) 103521}, [\href{https://arxiv.org/abs/astro-ph/0211258}{{\tt
  astro-ph/0211258}}].

\bibitem{Jedamzik:2006xz}
K.~Jedamzik, \emph{{Big bang nucleosynthesis constraints on hadronically and
  electromagnetically decaying relic neutral particles}},
  \href{http://dx.doi.org/10.1103/PhysRevD.74.103509}{\emph{Phys. Rev.} {\bf
  D74} (2006) 103509}, [\href{https://arxiv.org/abs/hep-ph/0604251}{{\tt
  hep-ph/0604251}}].

\bibitem{Poulin:2015opa}
V.~Poulin and P.~D. Serpico, \emph{{Nonuniversal BBN bounds on
  electromagnetically decaying particles}},
  \href{http://dx.doi.org/10.1103/PhysRevD.91.103007}{\emph{Phys. Rev.} {\bf
  D91} (2015) 103007}, [\href{https://arxiv.org/abs/1503.04852}{{\tt
  1503.04852}}].

\bibitem{Hufnagel:2018bjp}
M.~Hufnagel, K.~Schmidt-Hoberg and S.~Wild, \emph{{BBN constraints on MeV-scale
  dark sectors. Part II. Electromagnetic decays}},
  \href{http://dx.doi.org/10.1088/1475-7516/2018/11/032}{\emph{JCAP} {\bf 1811}
  (2018) 032}, [\href{https://arxiv.org/abs/1808.09324}{{\tt 1808.09324}}].

\bibitem{Forestell:2018txr}
L.~Forestell, D.~E. Morrissey and G.~White, \emph{{Limits from BBN on Light
  Electromagnetic Decays}},
  \href{http://dx.doi.org/10.1007/JHEP01(2019)074}{\emph{JHEP} {\bf 01} (2019)
  074}, [\href{https://arxiv.org/abs/1809.01179}{{\tt 1809.01179}}].

\bibitem{Cyburt:2015mya}
R.~H. Cyburt, B.~D. Fields, K.~A. Olive and T.-H. Yeh, \emph{{Big Bang
  Nucleosynthesis: 2015}},
  \href{http://dx.doi.org/10.1103/RevModPhys.88.015004}{\emph{Rev. Mod. Phys.}
  {\bf 88} (2016) 015004}, [\href{https://arxiv.org/abs/1505.01076}{{\tt
  1505.01076}}].

\bibitem{Kolb:1986nf}
E.~W. Kolb, M.~S. Turner and T.~P. Walker, \emph{{The Effect of Interacting
  Particles on Primordial Nucleosynthesis}},
  \href{http://dx.doi.org/10.1103/PhysRevD.34.2197}{\emph{Phys. Rev.} {\bf D34}
  (1986) 2197}.

\bibitem{Serpico:2004nm}
P.~D. Serpico and G.~G. Raffelt, \emph{{MeV-mass dark matter and primordial
  nucleosynthesis}},
  \href{http://dx.doi.org/10.1103/PhysRevD.70.043526}{\emph{Phys. Rev.} {\bf
  D70} (2004) 043526}, [\href{https://arxiv.org/abs/astro-ph/0403417}{{\tt
  astro-ph/0403417}}].

\bibitem{Nollett:2013pwa}
K.~M. Nollett and G.~Steigman, \emph{{BBN And The CMB Constrain Light,
  Electromagnetically Coupled WIMPs}},
  \href{http://dx.doi.org/10.1103/PhysRevD.89.083508}{\emph{Phys. Rev.} {\bf
  D89} (2014) 083508}, [\href{https://arxiv.org/abs/1312.5725}{{\tt
  1312.5725}}].

\bibitem{Nollett:2014lwa}
K.~M. Nollett and G.~Steigman, \emph{{BBN And The CMB Constrain Neutrino
  Coupled Light WIMPs}},
  \href{http://dx.doi.org/10.1103/PhysRevD.91.083505}{\emph{Phys. Rev.} {\bf
  D91} (2015) 083505}, [\href{https://arxiv.org/abs/1411.6005}{{\tt
  1411.6005}}].

\bibitem{Depta:2019lbe}
P.~F. Depta, M.~Hufnagel, K.~Schmidt-Hoberg and S.~Wild, \emph{{BBN constraints
  on the annihilation of MeV-scale dark matter}},
  \href{http://dx.doi.org/10.1088/1475-7516/2019/04/029}{\emph{JCAP} {\bf 1904}
  (2019) 029}, [\href{https://arxiv.org/abs/1901.06944}{{\tt 1901.06944}}].

\bibitem{Slatyer:2015jla}
T.~R. Slatyer, \emph{{Indirect dark matter signatures in the cosmic dark ages.
  I. Generalizing the bound on s-wave dark matter annihilation from Planck
  results}}, \href{http://dx.doi.org/10.1103/PhysRevD.93.023527}{\emph{Phys.
  Rev.} {\bf D93} (2016) 023527}, [\href{https://arxiv.org/abs/1506.03811}{{\tt
  1506.03811}}].

\bibitem{Bjorken:1988as}
J.~D. Bjorken, S.~Ecklund, W.~R. Nelson, A.~Abashian, C.~Church, B.~Lu et~al.,
  \emph{{Search for Neutral Metastable Penetrating Particles Produced in the
  SLAC Beam Dump}},
  \href{http://dx.doi.org/10.1103/PhysRevD.38.3375}{\emph{Phys. Rev.} {\bf D38}
  (1988) 3375}.

\bibitem{Dolan:2017osp}
M.~J. Dolan, T.~Ferber, C.~Hearty, F.~Kahlhoefer and K.~Schmidt-Hoberg,
  \emph{{Revised constraints and Belle II sensitivity for visible and invisible
  axion-like particles}},
  \href{http://dx.doi.org/10.1007/JHEP12(2017)094}{\emph{JHEP} {\bf 12} (2017)
  094}, [\href{https://arxiv.org/abs/1709.00009}{{\tt 1709.00009}}].

\bibitem{Alekhin:2015byh}
S.~Alekhin et~al., \emph{{A facility to Search for Hidden Particles at the CERN
  SPS: the SHiP physics case}},
  \href{http://dx.doi.org/10.1088/0034-4885/79/12/124201}{\emph{Rept. Prog.
  Phys.} {\bf 79} (2016) 124201}, [\href{https://arxiv.org/abs/1504.04855}{{\tt
  1504.04855}}].

\bibitem{Feng:2018noy}
J.~L. Feng, I.~Galon, F.~Kling and S.~Trojanowski, \emph{{Axionlike particles
  at FASER: The LHC as a photon beam dump}},
  \href{http://dx.doi.org/10.1103/PhysRevD.98.055021}{\emph{Phys. Rev.} {\bf
  D98} (2018) 055021}, [\href{https://arxiv.org/abs/1806.02348}{{\tt
  1806.02348}}].

\bibitem{Berlin:2018pwi}
A.~Berlin, S.~Gori, P.~Schuster and N.~Toro, \emph{{Dark Sectors at the
  Fermilab SeaQuest Experiment}},
  \href{http://dx.doi.org/10.1103/PhysRevD.98.035011}{\emph{Phys. Rev.} {\bf
  D98} (2018) 035011}, [\href{https://arxiv.org/abs/1804.00661}{{\tt
  1804.00661}}].

\bibitem{Dobrich:2015jyk}
B.~Döbrich, J.~Jaeckel, F.~Kahlhoefer, A.~Ringwald and K.~Schmidt-Hoberg,
  \emph{{ALPtraum: ALP production in proton beam dump experiments}},
  \href{http://dx.doi.org/10.1007/JHEP02(2016)018}{\emph{JHEP} {\bf 02} (2016)
  018}, [\href{https://arxiv.org/abs/1512.03069}{{\tt 1512.03069}}].

\bibitem{Slatyer:2017sev}
T.~R. Slatyer, \emph{{Indirect Detection of Dark Matter}},  in
  \emph{{Proceedings, Theoretical Advanced Study Institute in Elementary
  Particle Physics : Anticipating the Next Discoveries in Particle Physics
  (TASI 2016): Boulder, CO, USA, June 6-July 1, 2016}}, pp.~297--353, 2018.
\newblock \href{https://arxiv.org/abs/1710.05137}{{\tt 1710.05137}}.
\newblock \href{http://dx.doi.org/10.1142/9789813233348_0005}{DOI}.

\bibitem{Navarro:1995iw}
J.~F. Navarro, C.~S. Frenk and S.~D.~M. White, \emph{{The Structure of cold
  dark matter halos}}, \href{http://dx.doi.org/10.1086/177173}{\emph{Astrophys.
  J.} {\bf 462} (1996) 563--575},
  [\href{https://arxiv.org/abs/astro-ph/9508025}{{\tt astro-ph/9508025}}].

\bibitem{Navarro:1996gj}
J.~F. Navarro, C.~S. Frenk and S.~D.~M. White, \emph{{A Universal density
  profile from hierarchical clustering}},
  \href{http://dx.doi.org/10.1086/304888}{\emph{Astrophys. J.} {\bf 490} (1997)
  493--508}, [\href{https://arxiv.org/abs/astro-ph/9611107}{{\tt
  astro-ph/9611107}}].

\bibitem{Essig:2013goa}
R.~Essig, E.~Kuflik, S.~D. McDermott, T.~Volansky and K.~M. Zurek,
  \emph{{Constraining Light Dark Matter with Diffuse X-Ray and Gamma-Ray
  Observations}}, \href{http://dx.doi.org/10.1007/JHEP11(2013)193}{\emph{JHEP}
  {\bf 11} (2013) 193}, [\href{https://arxiv.org/abs/1309.4091}{{\tt
  1309.4091}}].

\bibitem{Ibarra:2012dw}
A.~Ibarra, S.~Lopez~Gehler and M.~Pato, \emph{{Dark matter constraints from
  box-shaped gamma-ray features}},
  \href{http://dx.doi.org/10.1088/1475-7516/2012/07/043}{\emph{JCAP} {\bf 1207}
  (2012) 043}, [\href{https://arxiv.org/abs/1205.0007}{{\tt 1205.0007}}].

\bibitem{Boddy:2015efa}
K.~K. Boddy and J.~Kumar, \emph{{Indirect Detection of Dark Matter Using
  MeV-Range Gamma-Ray Telescopes}},
  \href{http://dx.doi.org/10.1103/PhysRevD.92.023533}{\emph{Phys. Rev.} {\bf
  D92} (2015) 023533}, [\href{https://arxiv.org/abs/1504.04024}{{\tt
  1504.04024}}].

\bibitem{Bringmann:2008kj}
T.~Bringmann, M.~Doro and M.~Fornasa, \emph{{Dark Matter signals from Draco and
  Willman 1: Prospects for MAGIC II and CTA}},
  \href{http://dx.doi.org/10.1088/1475-7516/2009/01/016}{\emph{JCAP} {\bf 0901}
  (2009) 016}, [\href{https://arxiv.org/abs/0809.2269}{{\tt 0809.2269}}].

\bibitem{Gruber:1999yr}
D.~E. Gruber, J.~L. Matteson, L.~E. Peterson and G.~V. Jung, \emph{{The
  spectrum of diffuse cosmic hard x-rays measured with heao-1}},
  \href{http://dx.doi.org/10.1086/307450}{\emph{Astrophys. J.} {\bf 520} (1999)
  124}, [\href{https://arxiv.org/abs/astro-ph/9903492}{{\tt
  astro-ph/9903492}}].

\bibitem{Bouchet:2008rp}
L.~Bouchet, E.~Jourdain, J.~P. Roques, A.~Strong, R.~Diehl, F.~Lebrun et~al.,
  \emph{{INTEGRAL SPI All-Sky View in Soft Gamma Rays: Study of Point Source
  and Galactic Diffuse Emissions}},
  \href{http://dx.doi.org/10.1086/529489}{\emph{Astrophys. J.} {\bf 679} (2008)
  1315}, [\href{https://arxiv.org/abs/0801.2086}{{\tt 0801.2086}}].

\bibitem{Weidenspointner:99}
G.~Weidenspointner, \emph{{The Origin of the Cosmic Gamma-Ray Background in the
  COMPTEL Energy Range}}.
\newblock PhD thesis, Technical University of Munich, Munich, Germany, 1999.

\bibitem{Kappadath:98}
S.~C. Kappadath, \emph{{Measurement of the Cosmic Diffuse Gamma-Ray Spectrum
  from 800 keV to 30 MeV}}.
\newblock PhD thesis, University of New Hampshire, 1998.

\bibitem{Strong:2004de}
A.~W. Strong, I.~V. Moskalenko and O.~Reimer, \emph{{Diffuse galactic continuum
  gamma rays. A Model compatible with EGRET data and cosmic-ray measurements}},
  \href{http://dx.doi.org/10.1086/423193}{\emph{Astrophys. J.} {\bf 613} (2004)
  962--976}, [\href{https://arxiv.org/abs/astro-ph/0406254}{{\tt
  astro-ph/0406254}}].

\bibitem{Ackermann:2012pya}
{\scshape Fermi-LAT} collaboration, M.~Ackermann et~al., \emph{{Fermi-LAT
  Observations of the Diffuse Gamma-Ray Emission: Implications for Cosmic Rays
  and the Interstellar Medium}},
  \href{http://dx.doi.org/10.1088/0004-637X/750/1/3}{\emph{Astrophys. J.} {\bf
  750} (2012) 3}, [\href{https://arxiv.org/abs/1202.4039}{{\tt 1202.4039}}].

\end{thebibliography}\endgroup

\end{document}